\documentclass[draft, english]{volcanica-template} 

\newcommand{\Title}{Fluorescence enhancement of single V2 centers in a 4H-SiC cavity antenna} 
\newcommand{\shortTitle}{Fluorescence enhancement of single V2 centers in a 4H-SiC cavity antenna} 
\newcommand{\Author}{J. Körber et al.\xspace} 
\author[{{\affiliation{1}}}]
{Jonathan K\"orber~\orcidaffil{0000-0002-7531-0295}\Email{jonathan.koerber@pi3.uni-stuttgart.de}\SharedAuthor}

\author[{{\affiliation{1},\affiliation{2},\affiliation{3}}}]
{Jonah Heiler~\orcidaffil{0009-0000-4621-4782}\samethanks}

\author[{{\affiliation{4}}}]
{Philipp Fuchs~\orcidaffil{0000-0002-9966-6093}}

\author[{{\affiliation{5}}}]
{Philipp Flad}

\author[{{\affiliation{1}}}]
{Erik Hesselmeier~\orcidaffil{0000-0002-4560-1745}}

\author[{{\affiliation{1}}}]
{Pierre Kuna}

\author[{{\affiliation{6}}}]
{Jawad Ul-Hassan~\orcidaffil{0000-0001-9537-2226}}

\author[{{\affiliation{7}}}]
{Wolfgang Knolle}

\author[{{\affiliation{4}}}]
{Christoph Becher~\orcidaffil{0000-0003-4645-6882}}

\author[{{\affiliation{1},\affiliation{2},\affiliation{3}}}]
{Florian Kaiser~\orcidaffil{0000-0002-5844-1779}
\Email{florian.kaiser@list.lu}}

\author[{{\affiliation{1},\affiliation{8}}}]
{Jörg Wrachtrup}

\affil[{{\affiliation{1}}}]{					
3rd Institute of Physics, University of Stuttgart, Allmandring 13, 70569 Stuttgart, Germany.
}
\affil[{{\affiliation{2}}}]{					
Materials Research and Technology (MRT) Department, Luxembourg Institute of Science and Technology (LIST), 4422 Belvaux, Luxembourg.}
\affil[{{\affiliation{3}}}]{					
Department of Physics and Materials Science, University of Luxembourg, 4422 Belvaux, Luxembourg}
\affil[{{\affiliation{4}}}]{					
Universität des Saarlandes, Fachrichtung Physik, Campus E2.6, 66123 Saarbrücken, Germany.
}
\affil[{{\affiliation{5}}}]{					
4th Physics Institute and Reseach Center SCoPE, University of Stuttgart, Pfaffenwaldring 57, 70569, Stuttgart, Germany.
}
\affil[{{\affiliation{6}}}]{					
Department of Physics, Chemistry and Biology, Linköping University, 581 83 Linköping, Sweden.
}
\affil[{{\affiliation{7}}}]{					
Leibniz-Institute of Surface Engineering (IOM), Permoserstraße 15, 04318 Leipzig, Germany.
}
\affil[{{\affiliation{8}}}]{                    
Max Planck Institute for Solid State Research, Heisenbersgtraße 1, 70569 Stuttgart, Germany.}

\begin{document}


\FrontMatter{\protect{
\noindent Solid state quantum emitters are a prime candidate in distributed quantum technologies since they inherently provide a spin-photon interface. An ongoing challenge in the field, however, is the low photon extraction due to the high refractive index of typical host materials.
This challenge can be overcome using photonic structures.
Here, we report the integration of V2 centers in a cavity-based optical antenna.
The structure consists of a silver-coated, \SI{135}{\nano\meter}-thin 4H-SiC membrane functioning as a planar cavity with a broadband resonance yielding a theoretical photon collection enhancement factor of \SI{\sim34}{}.
The planar geometry allows us to identify over 20 single V2 centers at room temperature with a mean (maximum) count rate enhancement factor of 9 (15).
Moreover, we observe 10 V2 centers with a mean absorption linewidth below \SI{80}{\mega\hertz} at cryogenic temperatures.
These results demonstrate a photon collection enhancement that is robust to the lateral emitter position.


}}[]{}

\section*{INTRODUCTION}\label{sec:introduction}
\begin{figure*}[ht]
\includegraphics[width=\textwidth]{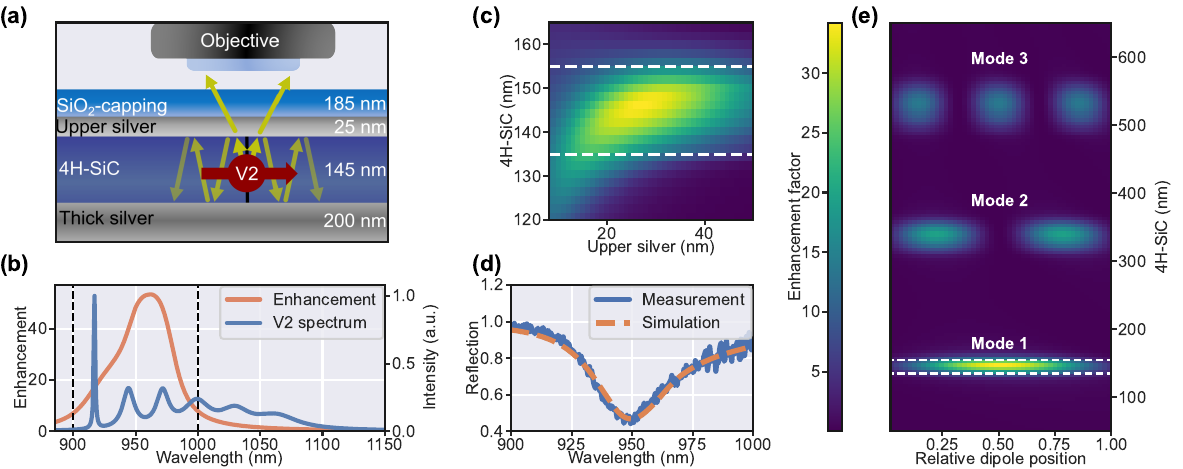}
\centering
\caption[]{\textbf{Optimization and measurement of the antenna structure}
\textbf{(a)} Schematic of the antenna structure.
\textbf{(b)} Theoretical enhancement of photon counts in the NA of the objective as a function of the wavelength (orange curve) and emission spectrum of the V2 center for reference (blue curve).
The black dashed lines indicate the spectral region where photons are collected in the experiment.
\textbf{(c)} Theoretical enhancement of photon counts between \SIrange{900}{1000}{\nano\meter} as a function of the thickness of the thin upper silver layer (x-axis) and the SiC thickness (y-axis) with a maximum of $\sim\SI{34}{}$.
White, dashed lines enclose the membrane thickness variation of the bright region in the fabricated structure.
\textbf{(d)} Measured reflectivity in the center of the fabricated antenna structure (blue curve) and modeled reflectivity using the transfer matrix method on a structure with a SiC thickness of \SI{137}{\nano\meter} and a thickness of the upper silver layer of \SI{22}{\nano\meter} (orange curve).
\textbf{(e)} Theoretical enhancement of photon counts between \SIrange{900}{1000}{\nano\meter} at optimum upper silver layer thickness of \SI{25}{\nano\meter} as a function of the 4H-SiC thickness (y-axis) and of the dipole position in the 4H-SiC. A relative dipole position of 0 corresponds to the dipole at the SiC-thick silver interface, 1 to the dipole at the SiC-upper silver interface. The simulation region is chosen to show the first three longitudinal modes of the cavity. White dashed lines enclose the membrane thickness variation of the bright region in the fabricated structure.
}
\label{Fig1}
\end{figure*}
Optically active spins in solid-state materials are promising candidates for applications in quantum technologies \cite{Awschalom2013,Awschalom2018,Atatuere2018}.
Besides the well-established diamond platform \cite{Jelezko2006,Doherty2013}, 4H-silicon carbide (4H-SiC) has gained growing attention in the field.
It hosts several promising emitters such as the silicon-vacancy center (V$_{\mathrm{Si}}$) \cite{Kraus2014,Widmann2014,Nagy2019}, the divacancy center (VV) \cite{Son2006,Christle2017,Li2022}, as well as the more recently investigated nitrogen-vacancy center \cite{VonBardeleben2015, Wang2020}, and the vanadium center \cite{Wolfowicz2020,Cilibrizzi2023}.
This has yielded a number of encouraging results with regards to quantum networks, such as single shot readout of the electron spin state, through charge state conversion using the VV center \cite{Anderson2022}, or through a nearby nuclear spin in case of the V$_{\mathrm{Si}}$ center \cite{Lai2024, Hesselmeier2024}, as well as spin-photon entanglement using the V$_{\mathrm{Si}}$ \cite{Fang2024}.

In addition, silicon carbide offers excellent nanofabrication capabilities.
This enables implementing various photonic structures that improve the rate of collected photons from emitters through their geometry, e.g. solid-immersion lenses (SIL) \cite{Sardi2020,Bekker2023}, nanopillars \cite{Radulaski2017}, and waveguides \cite{Babin2021,Krumrein2024}.
It also allows for the fabrication of more complex structures, such as photonic crystal cavities \cite{Bracher2017, Lukin2020} or disk resonators, \cite{Wang2021,Lukin2023} that can further boost the photon rate through Purcell enhancement. While nanophotonic cavities offer the highest enhancement among the above mentioned structures, they rely on complicated fabrication processes, tight restrictions on the emitter placement, and complex methods to keep the emitter on resonance with the structure.

Approaches like bullseye cavities \cite{Li2015,Hekmati2023}, plasmonic enhancement structures \cite{Shalaginov2019,Zhou2023}, and planar antenna structures \cite{Lee2011, Checcucci2017,Fuchs2021} represent an interesting middle ground.
They have shown a much higher photon count enhancement than SILs or nanopillars while offering relaxed restrictions on the emitter position. Additionally,  the operation conditions of such structures are less complicated since they work over a broader spectral range at the cost of a smaller count rate enhancement in comparison to nanophotonic cavities.

Inspired by previous results on tin vacancies in a diamond structure \cite{Fuchs2021}, we demonstrate the successful integration of single V2 centers (V$_{\mathrm{Si}}$ centers at the cubic lattice site in 4H-SiC) into a cavity-based optical antenna, consisting of a \SI{135}{\nano\meter}-thin 4H-SiC membrane with silver coatings on both sides.
We investigate 25 single V2 centers in a large region of the antenna ($\sim\SI{700}{\micro\meter\squared}$) at room temperature  and demonstrate an average fluorescence enhancement of factor 9 with a maximum of factor 15. 
Additionally, we show that the ODMR contrast of V2 centers in the structure is similar to the one of emitters in a bulk sample.
Further, we investigate 10 V2 centers in the antenna at cryogenic temperatures below \SI{10}{\kelvin} and show resonant absorption of the two optical transitions with mean linewidths below \SI{80}{\mega\hertz} and an inhomogeneous ZPL distribution of \SI{\pm170}{\giga\hertz}, which is comparable to our previous investigation of V2 centers in membranes \cite{Heiler2024}.
Lastly, we investigate the photon collection enhancement at low temperatures and find an increase in the photon count rate by a factor of 6.6 for the brightest emitter.

Compared to SILs or nanopillars, our structure offers a fluorescence enhancement over a much larger spatial region, only limited by the homogeneity of the membrane thickness.
Our results provide a promising route to strongly enhanced fluorescence with spin-selective excitation and thus could be used to boost readout rates in electron-nuclear spin experiments \cite{Lai2024,Hesselmeier2024}.
As shown in \cite{Fuchs2021}, our design could be adapted to a smaller resonance around the ZPL with an increased membrane thickness, i.e. with expectedly smaller absorption linewidths \cite{Heiler2024}, at the cost of a slightly reduced enhancement. Thus, it is also of high interest for applications in quantum networking, where a high ZPL rate and coherent optical lines are key requirements \cite{Northup2014, Ruf2021}.

\section*{RESULTS AND DISCUSSION}\label{sec:results}
\subsection*{Cavity antenna optimization and fabrication}
A schematic of the cavity antenna is shown in Figure \ref{Fig1}(a). The structure is based on a sub-micron-thin 4H-SiC membrane with silver coatings on both sides. Due to the high reflectance of the silver, the structure acts as a Fabry-Pérot cavity.
Choosing the thickness of the upper silver layer well below \SI{100}{\nano\meter} makes it transparent enough such that the cavity modes can be converted to free-propagating modes in the upper direction.
The lower silver layer is kept thick enough (several \SI{100}{\nano\meter}) to fully reflect the light.
This results in a higher photon emission rate and an increased directivity towards the collectable solid angle of an objective for the electric dipole radiation of a color center inside the SiC membrane.
Finally, the upper silver layer is capped by an SiO$_2$ layer to prevent fast oxidation of the silver at ambient conditions.
Its thickness is optimized to work as an anti-reflective coating~\cite{Fuchs2021}.
Note, that plasmonic resonances at the interfaces are a loss channel for this structure rather than the enhancement mechanism like in other approaches \cite{Shalaginov2019, Zhou2023}.

In contrast to the work of \cite{Fuchs2021}, the aim of our structure is to maximize the collectable photon rate within the phonon-side band (PSB) of the color center, in our case the V2 center.
Therefore, we optimize the structure to achieve a very broad resonance in the wavelength range of its PSB without including a second resonance that could be used for a more efficient off-resonant excitation.
This is enabled by tuning the membrane thickness to the first fundamental mode of the cavity, as depicted in Figure \ref{Fig1}(e).

For a given set of antenna layer thicknesses, the wavelength dependent emission enhancement over an emitter in bulk crystal is shown as the orange curve in Figure \ref{Fig1}(b).
We numerically optimize the thickness of the upper silver layer, the membrane, the SiO$_2$ capping layer, and the distance of the dipole from the lower silver layer to yield a maximum total enhancement \cite{Software2021} (see Supporting Information for details).
The total enhancement is calculated by weighting the wavelength dependent enhancement over the normalized fluorescence spectrum of the V2 center, shown in blue in Figure \ref{Fig1}(b).
Figure \ref{Fig1}(c) shows the total enhancement for the collection within a spectral band of \SIrange{900}{1000}{\nano\meter} as a function of the two most relevant thicknesses in the design, i.e., the upper silver layer and the membrane thickness. We choose this specific wavelength range, since our detector becomes much less efficient for wavelengths above \SI{1000}{\nano\meter} (see Supporting Information).
The structure yields a significant enhancement for membrane thicknesses between \SIrange{135}{155}{\nano\meter} and upper silver thicknesses between \SIrange{20}{40}{\nano\meter} with a maximum enhancement factor of \SI{\sim34}{}.
To study the influence of the dipole position within the thin SiC membrane, we simulate the enhancement factor as a function of the membrane thickness and the relative dipole position at the optimized upper silver layer thickness of \SI{25}{\nano\meter} in Figure \ref{Fig1}(e).
For the first cavity mode, the enhancement is very robust to a displacement of the dipole from its optimal position in the middle of the membrane.
It stays in between \SIrange{\sim 20}{34}{} for a relative dipole position in the membrane in between \SIrange{0.2}{0.8}{} from the lower silver-SiC interface.
The robustness of the enhancement on more parameters is shown in the Supporting Information.

The fabrication of the antenna structure starts with the steps described in our previous work \cite{Heiler2024} in order to create sub-µm thin 4H-SiC membranes with integrated V2-centers. With this, we create a sample with a total thickness of \SI{40}{\micro\meter} containing several subregions with a thickness below \SI{1}{\micro\meter}, which we refer to as membranes in the following.
In the last step of this process, we etch our sample down to a thickness of $\sim \SI{135}{\nano\meter}$ in the central part of one membrane (see Supporting Information for details on the membrane thickness), measured by a home-built white light interferometer (WLI).
The membrane is subsequently coated using electron beam evaporation with a target silver thickness of \SI{200}{\nano\meter} for the thick silver layer (full-reflective mirror) and \SI{25}{\nano\meter} for the thin silver layer (leaky mirror).
In the final step, the SiO$_2$ layer is sputtered on top of the thin silver layer.
To check the resonance of the fabricated structure, we measure its reflectivity spectrum after the coating steps using the WLI.
The result, normalised by a reference spectrum from a mirror, is depicted in Figure \ref{Fig1}(d) and shows a reflection dip at \SI{950}{\nano\meter} that covers a big part of the phonon side band of the V2 center.
We can model the measured data using a transfer matrix model and find a good agreement for an upper silver layer of \SI{22}{\nano\meter} and a SiC thickness of \SI{137}{\nano\meter}, confirming the earlier WLI measurement.

\subsection*{Fluorescence enhancement at room temperature}
\begin{figure}[!ht]
\includegraphics[width=\columnwidth]{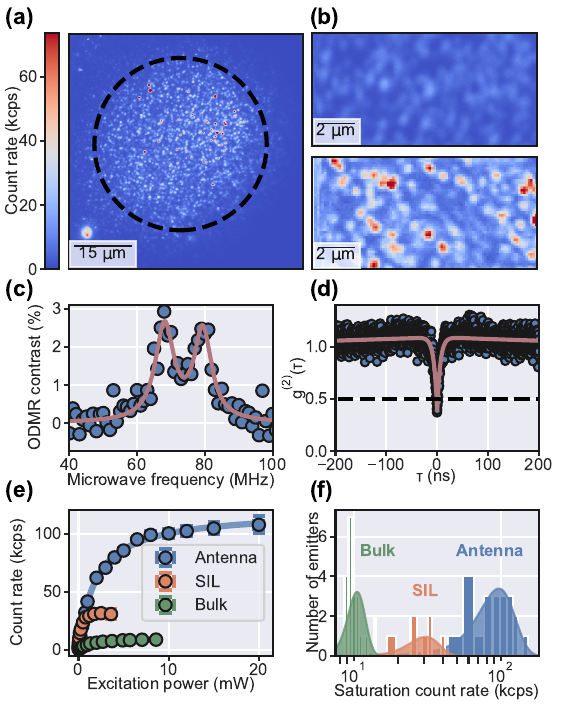}
\centering
\caption[]{\textbf{Room temperature fluorescence enhancement of single V2 centers.}
\textbf{(a)} Confocal xy-scan of the central part of the structure under excitation with  a \SI{785}{\nano\meter} laser.
The dashed circle encloses the region with increased fluorescence.
\textbf{(b)} Zoomed confocal xy-scan of a bulk reference sample (upper) and the structure (lower) at similar excitation levels.
\textbf{(c)} ODMR measurement from an emitter in the antenna sample (blue points).
The solid red line shows a double-Lorentzian fit on the measured data.
\textbf{(d)} Autocorrelation measurement from the same emitter as (c).
The red solid line shows a fit on the measured data points.
\textbf{(e)} Saturation measurements on the same emitter as (c) and (d) (blue) as well as each an emitter from a reference bulk (green) and SIL (orange) sample.
Solid lines are a fit to the data points to extract the saturation count rates.
\textbf{(f)} Histogram of the fitted saturation count rates for different emitters in the reference bulk (green), the SIL (orange) and the antenna (blue) sample.
The shaded regions show Gaussian distributions based on the mean value and standard deviation from the different saturation count rates.
The x-axis is scaled logarithmic for a better visibility of the three distributions.}
\label{Fig2}
\end{figure}

To test the enhancement of V2 fluorescence in our structure, we investigate the sample in a confocal laser scanning setup at room temperature (details in the Supporting Information).
At first, we measure fluorescence from the central part of the antenna and find a region with a diameter of $\sim \SI{40}{\micro\meter}$ around the membrane center that shows bright fluorescence, depicted as the black circle in Figure \ref{Fig2}(a).
The diameter of the enhanced region is determined by the thickness variation of the SiC membrane in this region.
We measured this variation to be $\sim \SI{20}{\nano\meter}$ (see Supporting Information for details), shown as white dashed lines in Figures \ref{Fig1}(c) and \ref{Fig1}(e), and see the corresponding enhancement drop outside of this region.
A zoom of the region reveals the presence of diffraction limited spots that appear significantly brighter than the spots in a scan taken on a bulk reference sample at similar excitation conditions, see Figure \ref{Fig2}(b).
To be comparable, V2 centers in the reference sample were created with electron irradiation at the same dose as for the antenna sample.

To confirm the presence of single V2 centers in the antenna sample, we continue with optically-detected magnetic resonance (ODMR) and autocorrelation measurements on the small, bright spots we find in a confocal scan of the sample (experimental details can be found in the Supporting Information).
Figures \ref{Fig2}(c) and \ref{Fig2}(d) show the results of these measurements on an exemplary V2 center in our structure.
We find an ODMR contrast of $2.5 \pm 0.4\,\%$, which is a typical value for the V2 center at room temperature \cite{Radulaski2017, Wang2017, Singh2023, Krumrein2024} and an anti-bunching of g$^{(2)}(\tau=\tau_0) = 0.38 \pm 0.01$, significantly below $0.5$ at zero time delay, without background correction, confirming that we are investigating a single V2-center.
We note that the typically observed single ODMR peak for the V2-center without an applied magnetic field is split by \SI{11}{\mega\hertz} into two peaks in our case.
Since this behavior is found for most V2-centers in different samples investigated on the setup with splittings between \SIrange{9}{13}{\mega\hertz} (see Supporting Information), we attribute it to a magnetization of parts of our setup.

To quantify the enhancement of the photon count rate of the exemplary emitter, we perform excitation power dependent measurements of the count rate on it and fit the saturation count rate to $119.3 \pm \SI{4.9}{\kilo\countpersec}$.
The same measurements are performed on a single V2 center in a reference bulk sample and on a single emitter in a SIL (see Supporting Information for details on the reference samples), yielding saturation counts of $10.1 \pm \SI{0.6}{\kilo\countpersec}$ and $33.9 \pm \SI{1.7}{\kilo\countpersec}$, as shown in Figure \ref{Fig2}(e).
These measurements show that the emitter in the antenna is enhanced by more than one order of magnitude compared to the bulk emitter and a factor of $\sim\SI{3.5}{}$ compared to the SIL emitter.

For better statistics on the enhancement factor, we investigate a higher number of emitters.
At first, we record ODMR for individual spots on the bulk and antenna sample to find V2 centers.
We set a threshold of at least $0.9\,\%$ contrast in the ODMR to use an emitter for further measurements, since a much lower contrast compared to the average indicates fluorescence from different defects, e.g., the V1 center, or crystal damage in the spot.
With this, we find that $18.8\,\%$ of the bright spots in the antenna are V2 centers which is lower compared to $28.8\,\%$ in the bulk reference sample.
We suspect the close proximity to the surface in our antenna sample yields a higher number of surface related defects in confocal scans, thus, decreases the relative number of V2 emitters.
However, with a mean density of \SI{0.29}{\per\micro\meter\cubed} we still find a high number of V2 centers in the antenna sample.
For a faster identification of V2 centers among the bright spots, we use a pre-selection based on the polarization of the emission (see Supporting Information for details).
In the following, we use 25 representative emitters in the antenna, 23 in the bulk, and 7 emitters in the SIL sample.

To ensure that only single emitters are included into our statistic, we measure the second order autocorrelation function. 
For the 25 antenna emitters, we find 14 (21) emitters with an autocorrelation dip below 0.5 before (after) applying background correction.
Due to the low brightness, we only test 7 of the 23 emitters in bulk for autocorrelation and all of them show a clear anti-bunching below 0.5 after background correction.
The 7 emitters from the SIL sample show anti-bunching below 0.5 after correction for the background as well.
The autocorrelation values for all emitters of the study can be found in the Supporting Information.

Subsequently, we measure excitation power dependent photon count rates for all single V2 centers to extract the saturation counts as described in the Supporting Information.
Figure \ref{Fig2}(f) shows a histogram of the saturation count rates for all emitters from the bulk, SIL, and antenna sample.
Including all emitters with an anti-bunching below 0.5 after applying background correction, we find an average enhancement of $\times$ (\SI{9.2\pm2.8}{}) for the antenna sample which is significantly higher than the average enhancement of $\times$ (\SI{2.9 \pm0.8}{}) from the SIL sample.
The maximum enhancement is $\times$ (\SI{15 \pm 2.9}) for the emitter with the highest saturation count rate in the antenna.

\subsection*{Absorption linewidths and inhomogeneous distribution of ZPLs under resonant excitation}
\begin{figure*}[ht]
\includegraphics[width=\textwidth]{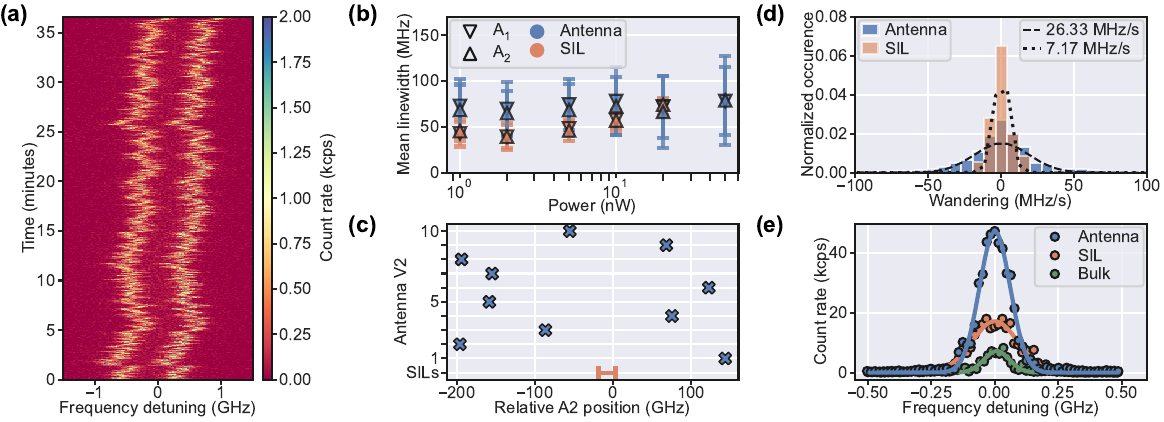}
\centering
\caption[]{
\textbf{Low temperature measurements.}
\textbf{(a)} Representative long time PLE scan at \SIrange{1}{2}{\nano\watt} resonant laser power of one V2 center in the structure without applying an additional microwave signal.
\textbf{(b)} Power dependence of the mean A$_1$ and A$_2$ linewidths for 10 V2 centers in the antenna structure and 1 emitter in a SIL.
The error bars correspond to one standard deviation of the linewidth data.
\textbf{(c)} Inhomogeneous spectral distribution of the mean A$_2$ transition peak of the 10 emitters in the antenna in relation to the mean transition peak of the SIL emitter and the range of frequency positions of 40 other emitters in SILs.
\textbf{(d)} Spectral wandering histogram of the emitters in the antenna structure (blue) and the SIL emitter (orange) together with a Gaussian function (black) based on the mean value and standard deviation of the respective data set.
The label for the Gaussian function shows the standard deviation of the data.
\textbf{(e)} Single PLE line for the brightest bulk (blue), SIL (orange) and antenna (green) emitter for highest possible excitation power before ionization. The solid lines are Gaussian fits to the data.
}
\label{Fig3}
\end{figure*}

Since applications based on  a spin–photon interface require an operation under cryogenic conditions, we investigate the optical properties of antenna-integrated V2 centers at temperatures below \SI{10}{\kelvin}.
To identify the V2 centers within diffraction limited spots, we measure their emission spectrum at off-resonant excitation in a low temperature confocal microscope setup (details see Supporting Information).
We investigate 32 representative emitters with a distinct emission peak in the range of \SIrange{916.2}{918.2}{\nano\meter}  by photoluminescence excitation (PLE) measurements.
Of those emitters, $47\,\%$ show two clearly distinguishable optical spin-conserving transitions A$_1$ and A$_2$, as expected for the V2 center \cite{Kraus2014}, while $15\,\%$ show a broad PLE and $38\,\%$ no PLE at all (see Supporting Information for details).

We note that we were able to measure PLE without an external microwave drive or a static magnetic field, which are typically necessary to counteract optical spin pumping \cite{Banks2019,Babin2021}. This indicates a spin-mixing of the ground state sub-levels which we suspect to originate from an increased amount of strain at low temperatures due to different thermal expansion of the layers.

At resonant excitation with low power of (\SIrange{1}{2}{\nano\watt}), we find emitters with a stable PLE over a long time scale, as shown for a representative emitter in Figure \ref{Fig3}(a) for half an hour.
We perform excitation power dependent PLE measurements to estimate the absorption linewidths of 10 arbitrarily selected emitters out of the emitters with distinguishable transitions.
Each emitter is measured for 10 single sweeps at up to six different excitation powers in between \SIrange{1}{50}{\nano\watt}.
The mean A$_1$ and A$_2$ linewidths for each power obtained by averaging the linewidth fit parameter from a double peak Gaussian fit of each single sweep are depicted in Figure \ref{Fig3}(b).
For comparison, the mean A$_1$ and A$_2$ linewidths of an emitter in a SIL sample are measured as well, using a spin-mixing microwave signal.
As expected due to the much larger distance of the emitter from the SiC surface, the linewidths from the SIL emitter lie below the respective antenna sample transition linewidths, except for the highest excitation power, where power broadening becomes significant for both structures.
In total, the average linewidth for each power for the antenna emitters lies below \SI{80}{\mega\hertz}.
Data on the power dependent linewidths of each individual antenna emitter can be found in the Supporting Information.

The mean spectral position of the A$_2$ peak for each measured emitter in the antenna compared to the respective position of the emitter in the SIL is depicted in Figure \ref{Fig3}(c).
The relative positions range from \SI{-196}{\giga\hertz} to \SI{144}{\giga\hertz}, spanning a total of \SI{340}{\giga\hertz}.
A smaller range of \SI{168}{\giga\hertz} was already observed in \SI{700}{\nano\meter} membranes \cite{Heiler2024}.
For comparison, the spectral range of A$_2$ peak positions in 40 measured SIL emitters from another sample, also shown in Figure \ref{Fig3}(c), amounts to \SI{22}{\giga\hertz}.
The bigger spread in thinner membrane structures can be explained by the increased influence of fabrication and strain on the integrated emitters.

Next, we quantify the spectral wandering as the positional change of the A$_2$ peak in consecutive single sweeps, to get a measure of the spectral stability over time.
This data is presented in form of histograms in Figure \ref{Fig3}(d).
As expected, the standard deviation of spectral wandering of emitters in the membrane with \SI{26}{\mega\hertz\per\second} is much bigger compared to the SIL emitter with \SI{7}{\mega\hertz\per\second} where the emitter is basically in bulk crystal environment.
This observation also aligns with previous work \cite{Heiler2024}.  

Finally, we investigate the count rate enhancement of the structure under resonant excitation.
To prevent a (partial) spin-polarization, which would lower the amount of collected photons, we scan two lasers separated by \SI{1}{\giga\hertz} simultaneously to drive the A$_1$ and A$_2$ transition (details in the Supporting Information). 
With this excitation scheme, we record power-dependent PLE for three emitters in the reference bulk, three emitters in the SIL, and 10 emitters in the antenna sample up to an excitation power where the emitter ionizes already after a single PLE scan most of the time.
We evaluate the photon count rate for each excitation power by averaging the amplitudes from Gaussian fits on the single PLE lines. 
However, we find that especially for the emitters in the antenna sample, the power dependent average count rates do not reach the typical non-linear regime of saturation since they ionize in almost every line before.
This effect results in saturation curves where no saturation power can be confidently extracted via a fit apart from 4 antenna emitters (see Supporting Information). 

The extracted average enhancement is $\times \left( \SI{2.2 \pm 0.5}{} \right)$ for the SIL and $\times \left( \SI{6.2 \pm 3.9}{} \right)$ for the antenna sample. To get an estimate of the maximum enhancement, we compare the brightest single PLE line with no sign of ionization from the bulk, SIL and antenna emitters in Figure \ref{Fig3}(e).
We extract an enhancement of $\times \left(\SI{2.4 \pm 0.1}{}\right)$ for the brightest SIL and $\times \left(\SI{6.6 \pm 0.3}{}\right)$ for the brightest antenna emitter.
The average enhancement in the antenna at low temperatures is $67\,\%$ of the enhancement extracted at room temperature.
This decrease is expected due to the higher detection efficiency at the low temperature setup for wavelengths above \SI{1000}{\nano\meter} where the structure shows a smaller enhancement (see Supporting Information for details).

\subsection*{Summary and Outlook}
In summary, we optimized and fabricated a cavity-based 4H-SiC antenna structure and demonstrated a collection count enhancement of factor 9 (15) in average (at most) at room temperature from integrated V2 centers.
At low temperatures, we found an average absorption linewidth below \SI{80}{\mega\hertz} and a collection count enhancement of factor 6.6 for the brightest checked emitter.
Further, we find a standard deviation in the spectral position over time of \SI{26}{\mega\hertz\per\second} which remains small enough to implement a resonance feedback system reported previously \cite{Babin2021}.

Compared to nanopillars, SILs, or nanobeam resonators that enhance emitters in a small spatial region, our structure offers the big advantage to enhance emitters in a two-dimensional plane, that is limited in size solely by the fabrication.
Thus, further work on the fabrication, i.e. creating all color centers in the optimal depth using ion-implantation methods \cite{Ohshima2018, Wang2019} and further optimizing the thickness homogeneity, can be used to significantly scale up the number of enhanced emitters in a single structure.
This simplifies the integration of defects with a low creation yield, e.g., SiC divacancies near stacking faults \cite{He2024}.
Moreover, if the structure is transferred to a system with emitters that have a broad inhomogeneous linewidth, e.g., optically active molecules \cite{Checcucci2017, Toninelli2021} or rare earth ion quantum emitters \cite{Thiel2011}, the big number of enhanced emitters eases finding emitters with matching emission lines, enabling two-emitter entanglement.
The versatile design of the cavity antenna also enables enhancing the ZPL using the spectrally smaller resonance of a thicker membrane and making the structure very appealing in the context of quantum communication where indistinguishable photons are key.

\section*{DATA AVAILABILITY}
The data supporting the presented findings are available at \url{https://doi.org/10.18419/darus-4262}.

\section*{CODE AVAILABILITY}
The measurement and evaluation codes used for this study are available from the corresponding author upon reasonable request. The optimization of our optical structure is based on the published python code \cite{Software2021}.

\section*{AUTHOR CONTRIBUTIONS}
J.K. and J.H. contributed equally to this work.
The project was conceived by J.K., J.H., P.Fu., C.B., and F.K. and supervised by F.K. and J.W.
The high-quality epilayer was grown by J.U.H. and the sample was electron irradiated by W.K.
J.K. and J.H. fabricated the SiC membrane and performed fabrication related measurements of the sample.
P.Fl. applied the thin-silver layer and the SiO$_2$ coating.
The optical measurements at room temperatures were conducted and analyzed by J.K.
The optical measurements at low temperatures were conducted by J.K., E.H., and P.K. and analyzed by J.H. and J.K.
The reference sample was fabricated by E.H.
The manuscript was written by J.K. and J.H.
All authors contributed to the manuscript.

\section*{COMPETING INTERESTS}
The authors declare no competing interests.

\section*{ACKNOWLEDGEMENTS}
We acknowledge fruitful discussions and experimental help from M. Krumrein, R. Stöhr, and A. Weible.

J.H., F.K., and J.W. acknowledge support from the European Commission for the Quantum Technology Flagship project QIA (Grant agreements No. 101080128, and 101102140).
P.K., F.K., J.U.H, and J.W. acknowledge support from the European Commission through the QuantERA project InQuRe (Grant agreements No. 731473, and 101017733).
P.K., F.K. and J.W. acknowledge the German ministry of education and research for the project InQuRe (BMBF, Grant agreement No. 16KIS1639K).
C.B., F.K. and J.W. further acknowledge the German ministry of education and research for the project QR.X (BMBF, Grant agreements No. 16KISQ001K and 16KISQ013), while J.W. also acknowledges support from the EU Horizon project SPINUS (project: 101135699).
F.K. and J.W. additionally acknowledge the Baden-Württemberg Stiftung for the project SPOC (Grant agreement No. QT-6).
F.K. acknowledges funding by the Luxembourg National Research Fund (FNR, project: 17792569) in addition.
J.U.H further acknowledges support from the Swedish Research Council under VR Grant No. 2020-05444 and Knut and Alice Wallenberg Foundation (Grant No. KAW 2018.0071).

\EndMatter
\end{document}


\title{Supporting Information to: \\ Fluorescence enhancement of single V2 centers in a 4H-SiC cavity antenna}

\author{Jonathan K\"orber}
\thanks{These authors contributed equally to this work}
\affiliation{
3rd Institute of Physics, University of Stuttgart, Pfaffenwaldring 57, 70569 Stuttgart, Germany.
}

\author{Jonah Heiler}
\thanks{These authors contributed equally to this work}
\affiliation{
3rd Institute of Physics, University of Stuttgart, Pfaffenwaldring 57, 70569 Stuttgart, Germany.
}
\affiliation{
Materials Research and Technology (MRT) Department, Luxembourg Institute of Science and Technology (LIST), 4422 Belvaux, Luxembourg.
}
\affiliation{
Department of Physics and Materials Science, University of Luxembourg, 4422 Belvaux, Luxembourg
}

\author{Philipp Fuchs}
\affiliation{
Universität des Saarlandes, Fachrichtung Physik, Campus E2.6, 66123 Saarbrücken, Germany.
}

\author{Philipp Flad}
\affiliation{
4th Physics Institute and Reseach Center SCoPE, University of Stuttgart, Pfaffenwaldring 57, 70569, Stuttgart, Germany.
}

\author{Erik Hesselmeier}
\affiliation{
3rd Institute of Physics, University of Stuttgart, Pfaffenwaldring 57, 70569 Stuttgart, Germany.
}

\author{Pierre Kuna}
\affiliation{
3rd Institute of Physics, University of Stuttgart, Pfaffenwaldring 57, 70569 Stuttgart, Germany.
}

\author{Jawad Ul-Hassan}
\affiliation{
Department of Physics, Chemistry and Biology, Linköping University, 581 83 Linköping, Sweden.
}

\author{Wolfgang Knolle}
\affiliation{
Leibniz-Institute of Surface Engineering (IOM), Permoserstraße 15, 04318 Leipzig, Germany.}

\author{Christoph Becher}
\affiliation{
Universität des Saarlandes, Fachrichtung Physik, Campus E2.6, 66123 Saarbrücken, Germany.
}

\author{Florian Kaiser}
\affiliation{
3rd Institute of Physics, University of Stuttgart, Pfaffenwaldring 57, 70569 Stuttgart, Germany.
}
\affiliation{
Materials Research and Technology (MRT) Department, Luxembourg Institute of Science and Technology (LIST), 4422 Belvaux, Luxembourg.
}
\affiliation{
Department of Physics and Materials Science, University of Luxembourg, 4422 Belvaux, Luxembourg
}

\author{J\"org Wrachtrup}
\affiliation{
3rd Institute of Physics, University of Stuttgart, Pfaffenwaldring 57, 70569 Stuttgart, Germany.
}
\affiliation{
Max Planck Institute for Solid State Research, Heisenbersgtraße 1, 70569 Stuttgart, Germany.}

\date{\today} 

\maketitle

\section{Methods} \label{sec:methods}

\subsection*{Structure optimization}
\begin{figure}[bt]
\centering
   \includegraphics[width=1\textwidth]{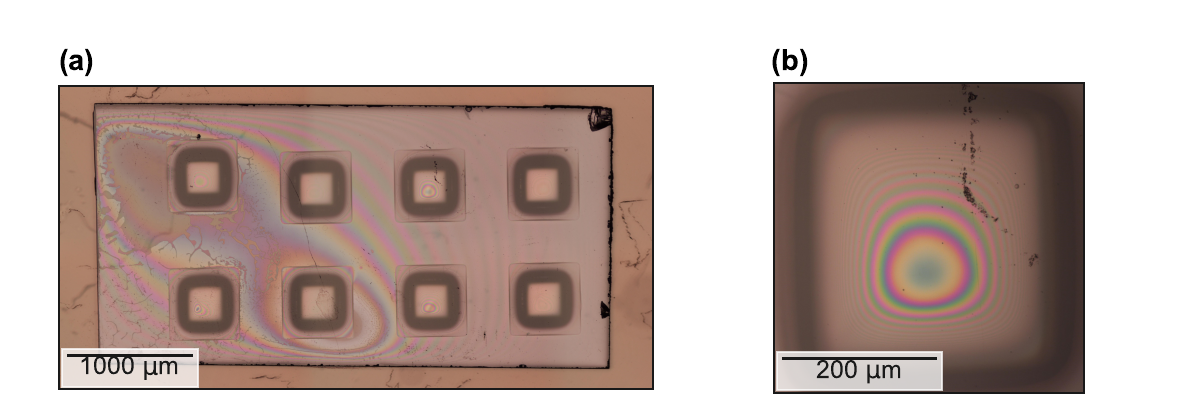}
\caption[]{
\textbf{Light microscope images of the sample after membrane fabrication (before applying the silver coatings).}
\textbf{(a)} Microscope image of the full sample after the membrane fabrication. The 8 visible squares are the membrane regions.
\textbf{(b)} Microscope image of the membrane that is used in the experiments. Color fringes arise due to a thickness gradient. The central part of the membrane is at a thickness of \SI{\sim 135}{\nano\meter}.
}
\label{SuppFigSample}
\end{figure}
\noindent To optimize the layer thicknesses of our design, we use a python code based on ref. \cite{Software2021}.
We simulate the power emitted by a dipole within our structure into the solid angle that can be collected by an objective with $\mathrm{NA}=0.9$, normalized by the power emitted from a dipole in bulk silicon carbide into the same solid angle of the air objective.
We choose a dipole orientation parallel to the interface since the sample is fabricated from a-plane silicon carbide, where the c-axis lies in the plane, and all V2 centers have a dipole parallel to the c-axis.
The resulting photon collection factor is calculated for different wavelengths to estimate an overall collection enhancement by summing up the values weighted by the normalized V2-spectrum at the respective wavelength.
For the optimization covering the full phonon side band, we do our calculations from \SIrange{900}{1150}{\nano\meter} with \SI{5}{\nano\meter} wavelength steps.
For the optimization in the other two spectral windows, we set the weighting factor to 0 for the wavelengths outside of the respective window.

\subsection*{Sample fabrication}
\noindent The SiC sample for this study is cut from an n-type, a-plane 4H-SiC wafer (\textit{Wolfspeed}) with a $\sim \SI{10}{\micro\meter}$-thick epilayer (free electron density of $\sim\SI{7e13}{\per\centi\meter\cubed}$) grown on top by chemical vapor deposition.
The creation of V2 centers and the fabrication of sub-micron thin membranes on our sample follows the steps from our previous work \cite{Heiler2024}.
An image of the sample after the final thinning containing several membranes at different central thickness is shown in Figure~\ref{SuppFigSample}.
The outer framing of the sample still has a thickness of around \SI{40}{\micro\meter}, such that the sample can be moved with tweezers for cleaning and the subsequent coating steps.
After identifying a membrane at the target thickness of $\sim\SI{135}{\nano\meter}$ (depicted in Figure \ref{SuppFigSample} (b)), we apply a $\sim\SI{200}{\nano\meter}$-thick silver layer by electron beam evaporation.
Subsequently, we flip our sample and apply the thin silver layer by electron beam evaporation as well.
As a last step, the SiO$_2$ layer is deposited to cover the thin silver layer, using ion-beam sputtering.
We note that we do not break the vacuum between the thin-silver evaporation and the SiO$_2$ sputtering since both processes use the same vacuum chamber.
In this way, the silver layer is covered even before its first contact with ambient conditions.

\subsection*{Room temperature confocal microscope setup}
\noindent Our room temperature confocal microscope setup consists of a continuous-wave (CW) excitation laser with $\lambda = \SI{785}{\nano\meter} (\textit{Cobolt 06-01, Hübner Photonics})$.
The laser is focused onto the sample with an objective (NA$=0.9$) and the fluorescence is separated by the excitation with a dichroic mirror (\textit{Semrock Di02-R830}). The collected fluorescence is then additionally filtered with a \SI{900}{\nano\meter} longpass filter (\textit{Thorlabs FELH900}), guided onto a polarizing beam splitter (PBS), coupled into two single-mode optical fibers (\textit{Thorlabs 780 HP}) and guided onto SNSPDs (\textit{PhotonSpot}).
To address different positions on the sample, we scan our objective by piezo-actuators with a maximum range of $\SI{100}{\micro\meter} \times \SI{100}{\micro\meter} \times \SI{25}{\micro\meter}$.

\subsection*{ODMR measurements}
\noindent To perform ODMR measurements on our sample at room temperature, we span a \SI{50}{\micro\meter}-thick copper wire across the sample with a distance of $\sim\SI{200}{\micro\meter}$ to the membrane center.
We use a signal generator (\textit{Rohde \& Schwarz, SMIQ03}) and a microwave amplifier (\textit{Mini Circuit LZY-22+}) to create a microwave sweep from typically \SIrange{30}{110}{\mega\hertz} with a power of typically \textbf{\SIrange{20}{24}{\dbmilliwatts}} sent through the wire after the amplifier.
During the sweep, we detect fluorescence while continuously exciting the color center off-resonantly at \SI{785}{\nano\meter} as a function of the microwave frequency in \SI{1}{\mega\hertz} steps with an integration time of \SI{50}{\milli\second} for each point.
Finally, we fit the acquired data using a double-Lorentzian function.

\subsection*{Autocorrelation measurements}
\noindent For measuring the autocorrelation of the fluorescence, we make use of a standard Hanbury-Brown and Twiss (HBT) interferometer.
To set an equal $50/50$ ratio of photon counts between the two detectors, we adjust a $\lambda/2$-waveplate in front of the PBS of our confocal setup, accordingly.
For the correlation measurement, we use a timetagger (\textit{Swabian  Instruments}) and bin the detected correlations typically in \SI{200}{\pico\second} intervals.
To analyze the data, we fit the autocorrelation measurements using a three level autocorrelation function for N single photon emitters \cite{Fuchs2015} 
\begin{equation*}
    \frac{1}{N}\cdot \left[ 1 - (1+a) \cdot \mathrm{e}^{-\left| \tau - \tau_0 \right| / \tau_1} + a \cdot \mathrm{e}^{-\left| \tau - \tau_0 \right| / \tau_2} \right] + \frac{N-1}{N}
\end{equation*}
where $a$, $\tau_0$ and $\tau_1$ are excitation power and intersystem crossing dependent parameters.
From here, we extract the single photon purity as g$^{(2)}(\tau_0) = (N-1)/N$.
For background correction, we use the form \cite{Fishman2023} 
\begin{equation*}
    \mathrm{g}^{(2)}_{\mathrm{corr}} = \frac{1}{\rho^2} \left[ \mathrm{g}^{(2)}_{\mathrm{meas}} - \left(1 - \rho^2 \right) \right]
\end{equation*}
with the background and emitter intensity dependent parameter $\rho = I_{\mathrm{em}}/(I_{\mathrm{em}} + I_{\mathrm{bg}})$.

\subsection*{Saturation measurements}
\noindent To extract the saturation count rate from single emitters, we measure the detected photon counts for different excitation power with the \SI{785}{\nano\meter} laser.
The given values of the excitation power are always measured directly before the objective.
Each data point in a saturation measurement is the average of 100 measured count rates with each \SI{100}{\milli\second} integration time.
As an error for each data point we use the standard deviation of the 100 measurements.
To extract the saturation behavior, we fit the standard saturation curve with a linear term to include the background on each spot
\begin{equation}
    I_{\mathrm{em}} (P) = \frac{I_{\mathrm{sat}} \cdot P}{P + P_{\mathrm{exc}}} + b \cdot P. 
\end{equation}
using an orthogonal distance regression algorithm with python.
We decide to include the background in the fit, since we find the background measured in close vicinity of individual spots strongly fluctuating for different regions even for regions very close to a specific emitter and thus does not give an accurate description of the background of each emitter.
After the fit, we estimate the background corrected count values by correcting the measurements with the fit results.

\subsection*{Low temperature confocal microscope setup}
\noindent For low temperature measurement, the sample is mounted inside a closed-cycle helium cooled cryostat (\textit{Montana Instruments}) at $\sim\SI{8}{\kelvin}$ and emitters are addressed using a confocal setup with a NA\SI{=0.9}{}, WD\SI{=1}{\milli\meter}, $100\times$ objective (\textit{Zeiss EC Epiplan-Neofluar}).
The emitters are off-resonantly excited using a \SI{728}{\nano\meter} laser (\textit{Toptica iBeam-Smart-CD}) and the spectra are measured using a spectrometer (\textit{Ocean Optics NIRQuest}).
The frequency of the resonant excitation laser (\textit{Toptica DL pro}) is recorded by a wavemeter (\textit{HighFinesse Angstrom WS7-60}) and the resulting photons pass two tunable long pass filters (\textit{Semrock TLP01-995}) with their edge tuned to $\sim\SI{925}{\nano\meter}$ for PSB detection before entering a single-mode fiber (\textit{Thorlabs 1060 XP}) connected to the SNSPDs (\textit{Photon Spot}).

\subsection*{PLE measurements}
\noindent For the PLE measurements, we use the tunable laser in our low temperature confocal microscope setup.
We apply a voltage ramp to the laser using a \textit{National Instruments} data acquisition card (NIDAQ) and simultaneously record the laser frequency using a wavemeter (\textit{HighFinesse Angstrom WS7-60}) as well as the photon counts in the PSB with our SNSPDs.
For measurements on the reference sample, we drive the ground state spin transition continuously during PLE by applying a microwave signal (\SIrange{12}{15}{\dbmilliwatts} input power), guided through a \SI{50}{\micro\meter}-thick wire to the emitter.
For our measurements where we simultaneously scan two laser lines, we split the resonant laser into two paths and shift the optical frequency of both parts to a relative difference of \SI{1}{\giga\hertz} using two acousto-optic modulators (\textit{Gooch\&Housego}).

\subsection*{Evaluation of PLE linewidths and spectral wandering}
\noindent The PLE count rate is recorded over the voltage ramp that is applied to tune the resonant laser.
In the first step of the evaluation, the voltage data is converted to frequencies using the data from the wavemeter for the respective ramp.
A linear ramp between the minimum and maximum frequency recorded by the wavemeter is assumed for each PLE line which is in good agreement with the experimental data.
Subsequently, the peaks in each PLE line are fitted using two Gaussian functions.
To avoid corrupted fit data from lines with strong spectral jumping or ionization in between, a few constraints on the fit are employed.
These constraints include peak positions in the scanning range, no negative amplitudes and offset, a linewidth bigger than one frequency step, an amplitude ratio below 10, a peak separation deviation below \SI{40}{\percent}, and a general fit quality, determined by the r2 value, bigger than one of two thresholds 0.46 (0.7923).
The power dependent mean linewidths are directly extracted from the fit data of all respective PLE measurement.
The spectral wandering is calculated as the difference of the fitted A$_2$ peak position in two consecutive lines divided by the measurement time of one line.

\section{Detection efficiency of the room temperature setup} \label{sec:DetEff}
\begin{figure}[tb]
\centering
   \includegraphics[width=1\textwidth]{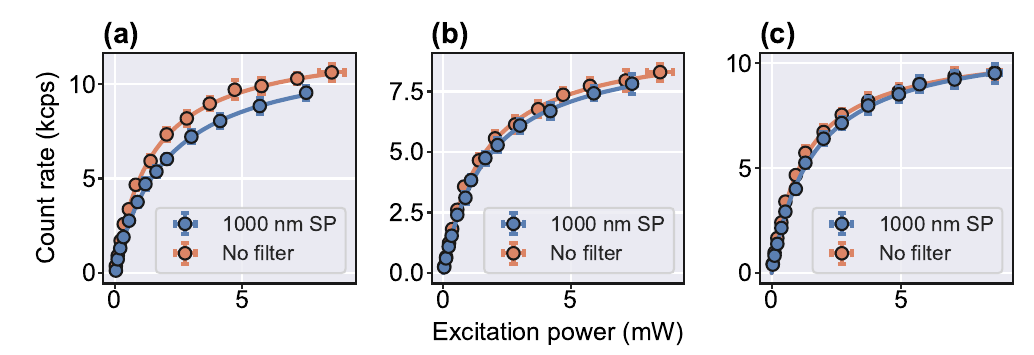}
\caption[]{
\textbf{Saturation of 3 exemplary emitters at different filtering.}
Count rate under off-resonant excitation at the room temperature setup for three representative V2 centers from \textbf{(a)} to \textbf{(c)}.
The orange data sets are taken without filtering of the emission while for the blue data sets a \SI{1000}{\nano\meter} shortpass filter was placed in the detection path.
Solid lines show saturation fits on the datasets.
}
\label{SuppFigDetection}
\end{figure}
\noindent As stated in the main manuscript, the detection efficiency of our SNSPDs at the room temperature setup drops significantly for wavelengths larger than \SI{1000}{\nano\meter} according to the spec sheets.
To measure the influence on the detected photon count rate, we performed the saturation measurements of all emitters on the bulk reference samples for a second time with an additional \SI{1000}{\nano\meter} shortpass filter (\textit{Thorlabs FESH1000}) in the detection path.
The measured saturation of 3 representative emitters are shown together with fits in Figure \ref{SuppFigDetection}.
Although the V2 emission ranges from $\sim$ \SIrange{917}{1100}{\nano\meter} with a significant amount above \SI{1000}{\nano\meter} (see Figure 1(b) of the main manuscript), the saturation measurements only show a very small difference between the two filter settings.
For all V2 emitters measured in the statistics of the main manuscript, the average saturation count rate yields \SI{10.5 \pm 1.4}{\kilo\countpersec} without an emission filtering (as used in the main manuscript) and \SI{10.1 \pm 1.7}{\kilo\countpersec} with the \SI{1000}{\nano\meter} shortpass filtering.
Therefore, the enhancement simulations in the main manuscript are always done for the spectral window of \SIrange{900}{1000}{\nano\meter}, comparable to the room temperature studies.
For completeness and to understand the enhancement decrease at the low temperature setup, where a different SNSPD with a higher detection efficiency for wavelength above \SI{1000}{\nano\meter} is used, we repeated the simulations for the different spectral windows in section \ref{sec:suppSweeps}.

\newpage
\section{Parametersweeps for the four relevant layer thicknesses} \label{sec:suppSweeps}
\begin{figure}[tb]
\centering
   \includegraphics[width=1\textwidth]{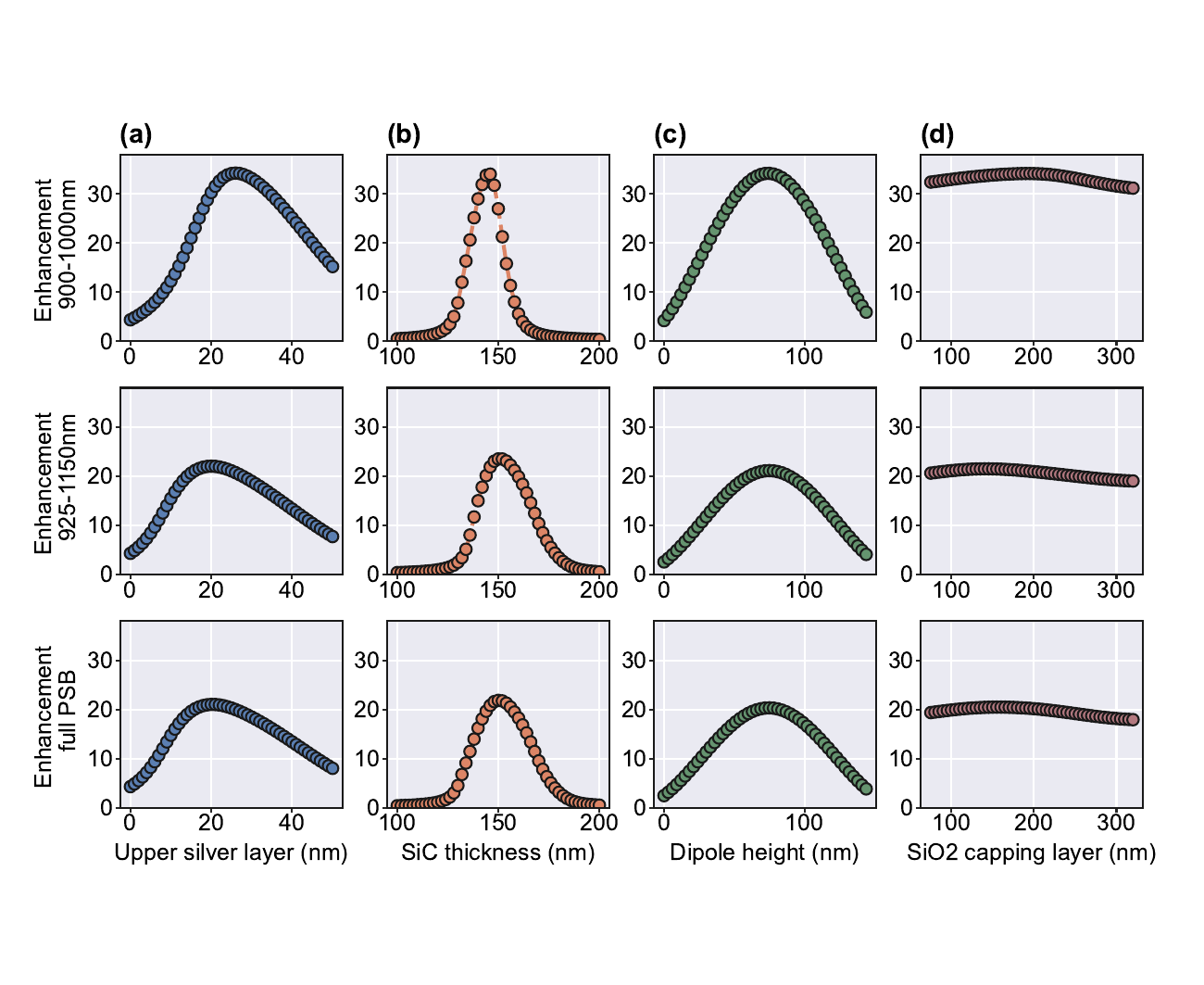}
\caption[]{
\textbf{Enhancement as a function of the four relevant parameters.}
Simulated enhancement for different upper silver layer thickness \textbf{(a)}, SiC thickness \textbf{(b)}, height of the dipole with respect to the lower SiC-Ag interface \textbf{(c)}, and thickness of the SiO$_2$ capping layer \textbf{(d)}.
For the upper row, the enhancement was simulated for collection of V2-emission in a spectral range from \SIrange{900}{1000}{\nano\meter} (as used in the RT measurements), for the central row in a spectral range from \SIrange{925}{1150}{\nano\meter} (as used in the LT measurements), and for the lower row for collection of the full PSB from \SIrange{900}{1150}{\nano\meter}.
}
\label{SuppFigSimulation}
\end{figure}

\noindent To evaluate the effect of fabrication errors and different spectral detection windows to the enhancement of our structure, we extract the enhancement from the simulations explained in the methods section of the main text for different parameter sweeps.
The enhancement is defined as the collected photon count rate of an emitter inside the structure divided by the count rate of a bulk emitter.
Figure \ref{SuppFigSimulation}(a) to (d), respectively, show sweeps of the thickness of the upper silver layer, the thickness of the silicon carbide membrane, the height of the dipole emitter inside the membrane, and the thickness of the silica capping layer.
For every parameter sweep, the three other parameters are kept at their optimum value.
The different spectral detection windows are chosen to be \SIrange{900}{1000}{\nano\meter}, as used in the room temperature measurements, \SIrange{925}{1150}{\nano\meter}, as used in the low temperature measurements, and the full PSB \SIrange{900}{1150}{\nano\meter} for reference (see \ref{SuppFigSimulation} from upper to lower row). 

As stated in the main manuscript, variations in the silica layer do not change the photonic enhancement significantly and this layer is thus not critical in the fabrication.
The most critical parameter to the enhancement is the thickness of the silicon carbide membrane, since it changes the resonance frequency of the structure.
Here, the membrane has a thickness of $\sim \SI{135}{\nano\meter}$ at its central part, which is slightly below the optimum thickness.
However, this is very beneficial for our case, since the membrane shows a thickness inhomogeneity and becomes thicker away from the center.
The upper silver layer optimum is at a thickness of $\sim \SI{20}{\nano\meter}$ ($\sim \SI{25}{\nano\meter}$ for \SIrange{900}{1000}{\nano\meter}), which we hit well in the fabrication.
Since our sample is electron irradiated, V2 centers are created evenly distributed within the membrane and not only in the optimum dipole height which is roughly at the center of the membrane.
Thus, the expected average enhancement in our case are smaller than the optimum values obtained from the simulations. 

When we compare the enhancement for different spectral windows, the shape of the parameter variations remain almost unchanged but the overall enhancement changes.
This is mainly due to the fact that our resonance is narrower than the PSB, meaning that it does not yield an enhancement in the outer parts of the side band, thus, decreasing the average enhancement when measuring the full PSB.
For our room temperature measurements, we are not able to measure the full PSB, since the efficiency of the detectors drops significantly for wavelengths above \SI{1000}{\nano\meter}.
When using the structure with a detection system that covers the full spectral bandwidth of the PSB, the enhancement is lowered by a factor of $\sim 0.65$, yielding an expected enhancement of $\times \, 22$ for the optimum case instead of $\times \,34$.
In our low temperature measurements, we use different detectors with a much higher detection efficiency above \SI{1000}{\nano\meter} but we limit our detection window to \SIrange{925}{1150}{\nano\meter} since the ZPL needs to be filtered in emission to avoid laser counts from the resonant excitation.
As shown in the last row of Figure \ref{sec:suppSweeps}, the enhancement in this case is lowered by a factor of $\sim 0.7$.

\section{Membrane thickness of the region with increased fluorescence}\label{sec:suppConfocalLateral}
\begin{figure}[tb]
\centering
   \includegraphics[width=0.4\textwidth]{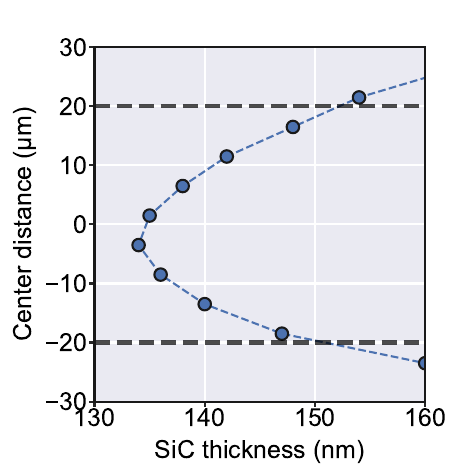}
\caption[]{
\textbf{Thickness in the central part of the membrane.}
Membrane thickness for different vertical distance from the membrane center.
The data is measured with the home-built WLI before the silver coating of the membrane.
Black dashed lines enclose the vertical region around the membrane center with bright fluorescence from the main manuscript.
The blue dashed line is a guide to the eye.
}
\label{SuppConfocalLateral}
\end{figure}

\noindent In our room temperature confocal measurements, we have found a big, circular area with bright fluorescence in the center of the antenna, shown in Figure 2(a) of the main manuscript.
The diameter of this region is around \SI{40}{\micro\meter}.
Measurements with our home-built WLI, taken before the silver coating, are depicted in Figure \ref{SuppConfocalLateral}.
They show a minimum thickness of \SI{134}{\nano\meter} and a maximum thickness of \SI{153}{\nano\meter}.
As discussed in the main text, the simulation shows the maximum enhancement for a membrane thickness between \SIrange{135}{155}{\nano\meter}, thus, matching our experiments.

\section{Reference bulk and SIL samples for this study} \label{sec:suppReference}
\begin{figure}[ht]
\centering
   \includegraphics[width=1\textwidth]{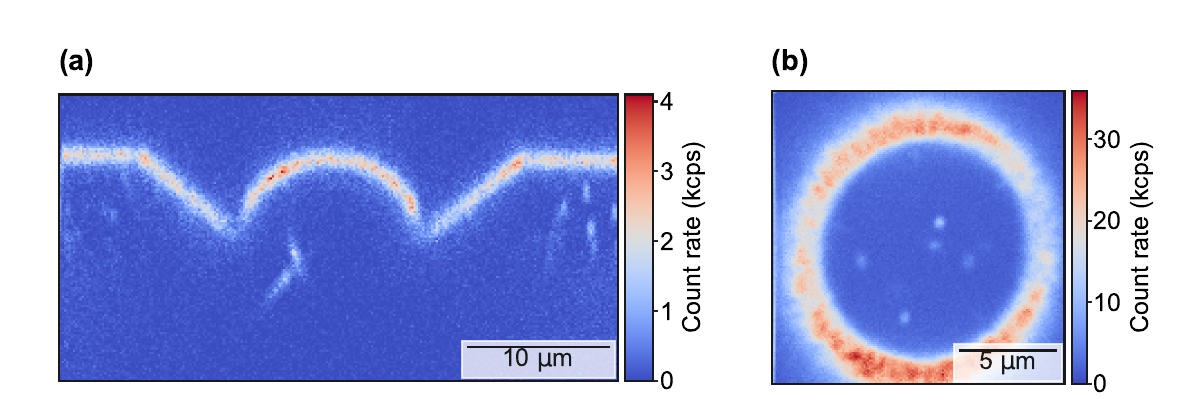}
\caption[]{
\textbf{Confocal scan of a SIL on the reference sample.}
\textbf{(a)} XZ-scan of a SIL fabricated on the reference sample under \SI{100}{\micro\watt} excitation with the \SI{785}{\nano\meter} laser.
The y-position is fixed to the center of the SIL.
The cross-section of the SIL surface is visible in the upper part of the image.
In the lower part one can see single emitters in the SIL as well as in the bulk part next to the SIL.
\textbf{(b)} XY-scan of the SIL with \SI{90}{\micro\watt} excitation (\SI{785}{\nano\meter}) at a z position a couple of \SI{}{\micro\meter} deep in the SIL.
Several, single emitters are visible in the scan. 
}
\label{SuppFigReference}
\end{figure}
\noindent The samples used for the reference measurements are fabricated from the same starting material, i.e., n-type 4H-SiC wafer with a $\sim \SI{10}{\micro\meter}$-thick epilayer.
The reference sample from Figure 2(b) of the main manuscript has been irradiated with the same dose of \SI{5}{\kilo\gray} as the main sample.
The second reference sample, used for the bulk and SIL reference measurement for the room temperature enhancement, has been irradiated with a dose of \SI{2}{\kilo\gray}.
Eight SILs with a diameter of $\SI{10}{\micro\meter}$ were fabricated using focused-ion-beam (FIB) milling with a gallium beam.
Figure \ref{SuppFigReference} shows confocal scans of such a created SIL.
For the bulk measurements, a region $\sim \SI{50}{\micro\meter}$ next to a SIL was used.
The last reference sample, used for measuring the spectral distribution of the A2 position in 40 SIL emitters, is made of the same starting material but has been irradiated with a dose of \SI{20}{\kilo\gray} before the SIL fabrication.

\section{ODMR for different samples at the room temperature setup} \label{sec:ODMR}
\begin{figure}[tb]
\centering
   \includegraphics[width=1\textwidth]{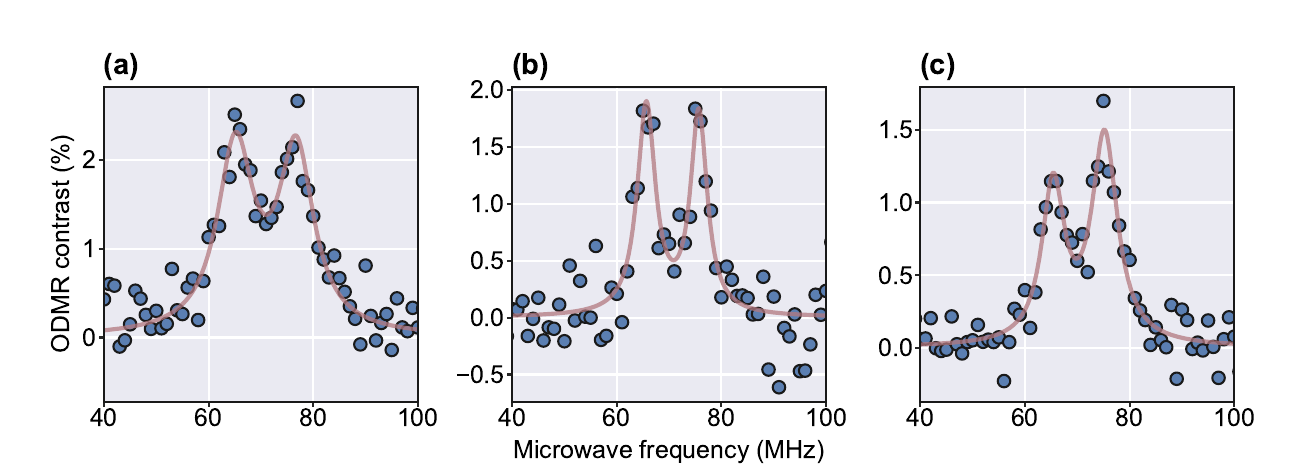}
\caption[]{
\textbf{ODMR measurements on single V2 emitters of three different samples.}
ODMR measurements of a single V2 emitter on the antenna sample \textbf{(a)}, the reference SIL sample \textbf{(b)}, and a third, different sample \textbf{(c)}.
The solid red lines show double-Lorentzian fits to the data.
}
\label{SuppODMR}
\end{figure}

\noindent As stated in the main manuscript, our room temperature ODMR measurements on V2 centers show two peaks.
To exclude effects introduced by our optical structure as the origin, we measured ODMR for V2 centers on different samples in addition.
Figure \ref{SuppODMR} shows ODMR measurements for three different samples, each with two peaks centered around \SI{70}{\mega\hertz}.
The splittings extracted from the fits are $\SI{11.6 \pm 0.4}{\mega\hertz}$, $\SI{10.0 \pm 0.3}{\mega\hertz}$, and $\SI{9.8 \pm 0.3}{\mega\hertz}$ from left to right.
Since the splittings are in the same range (between \SIrange{9}{13}{\mega\hertz} for all the investigated emitters) we conclude that our setup has a magnetized part very close to the sample (most likely the objective) which creates the splitting by a magnetic field of $\sim \SIrange{3}{5}{\gauss}$ parallel to the c-axis \cite{Kraus2014-2}.

\section{Polarization based preselection of bright confocal spots} \label{sec:Preselection}
\begin{figure}[tb]
\centering
   \includegraphics[width=1\textwidth]{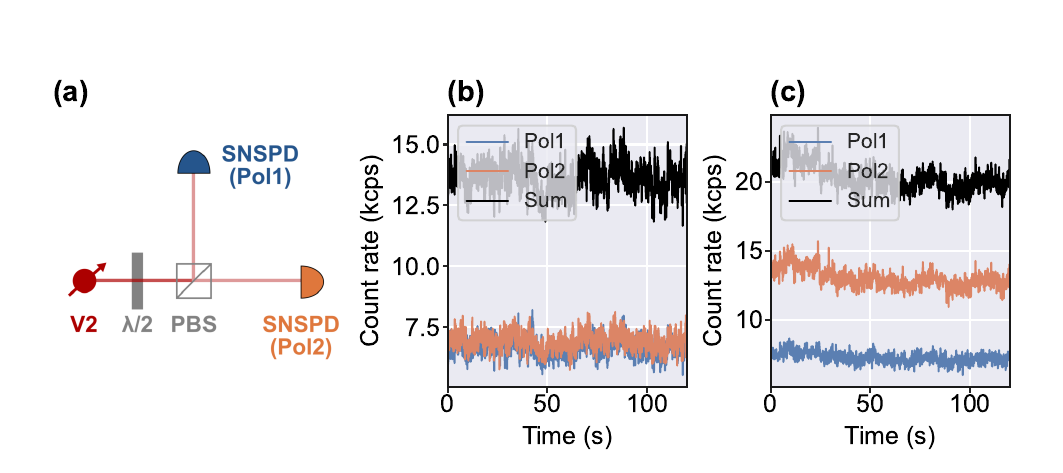}
\caption[]{
\textbf{Polarization of the collected emission from bright spots.}
\textbf{(a)} Schematic of the detection path in our room temperature confocal setup.
The emitted light passes a $\lambda$/2-waveplate and is directed onto a polarizing beamsplitter (PBS), that splits the beam into two paths, both sent to an SNSPD.
\textbf{(b)} Recorded time trace of photon count rates of a V2 emitter.
The orange (blue) curve shows the time trace of the SNSPD for horizontal (vertical) polarization and the black curve shows the sum of the two.
\textbf{(c)} Recorded photon count rate trace of a bright spot that is not a V2 emitter with the same $\lambda$/2 setting as in (b).
The polarization of the emission has significantly changed compared to (b).
}
\label{SuppFigPol}
\end{figure}

\noindent To find a high number of V2 centers within the bright spots in our antenna sample without the need of doing an ODMR measurement for all of them, we make use of our polarization dependent collection path.
As depicted in Figure \ref{SuppFigPol}(a), we place a $\lambda$/2-waveplate in front of a polarizing beam splitter (PBS) in our detection beam and send the two outputs of the PBS on an SNSPD.
We originally implemented this setup to be able to tune a 50/50 illumination of both SNSPDS for the autocorrelation measurements.
However, we can also use it to detect candidates for V$_{\mathrm{Si}}$ centers in 4H-SiC since their dipole emission is always oriented along the c-axis of the crystal \cite{Janzen2009}.
To do so, we use the emission from a spot that shows a very clear ODMR peak as a calibration and rotate the $\lambda$/2-waveplate to set an almost perfect 50/50 ratio on both detectors.
This setting is then fixed for further measurements.
When the dominant emission contribution of a bright spot on a different position originates from a V$_{\mathrm{Si}}$, the polarization and thus the count rate ratio of both detectors should not change, as shown in a time trace measurement of photon count rates for $100 \, \mathrm{ms}$ integration time in Figure \ref{SuppFigPol}(b).
For other spots, however, this ratio can differ quite significantly from the initially set 50/50 ratio, as shown in Figure \ref{SuppFigPol}(c), indicating that the collected fluorescence does not come from a V$_{\mathrm{Si}}$ center.
To quantify this effect, we define the polarization deviation $\Delta $Pol using the measured count rate of detector 1 (Pol1) and detector 2 (Pol2) for N time traces
\begin{equation*}
    \Delta \mathrm{Pol} = \frac{100}{N}\sum_{N} \left( \frac{1}{2} - \frac{\mathrm{Pol1}}{\mathrm{Pol1} + \mathrm{Pol2}} \right).
\end{equation*}
The polarization deviation of the two shown examples are $\Delta \mathrm{Pol} = \SI{0.91}{\percent}$ for Figure \ref{SuppFigPol}(b) and $\Delta \mathrm{Pol} = \SI{14.16}{\percent}$ for Figure \ref{SuppFigPol}(c).
A the perfect value of $\Delta \mathrm{Pol} = \SI{0}{\percent}$ cannot be reached in the experiment since the $\lambda$/2-plate being turned by hand and due to the influence of shot noise in the photon detection.
We use this preselection and only consider spots with $\Delta \mathrm{Pol} < \SI{1.5}{\percent}$ for further investigations with ODMR measurements in the room temperature study.

Using this preselection, we increase the relative amount of V2 centers from the $18.8 \%$ in the antenna sample to $45 \%$. Note that this is in good agreement with the expectation of roughly $50 \%$, as both the V2 and the V1 center are found with our preselection method and we expect to create those color centers with similar probability during the electron irradiation.

\newpage
\section{Autocorrelation values for all measured emitters} \label{sec:autocorrelation}
\begin{figure}[tb]
\centering
   \includegraphics[width=0.8\textwidth]{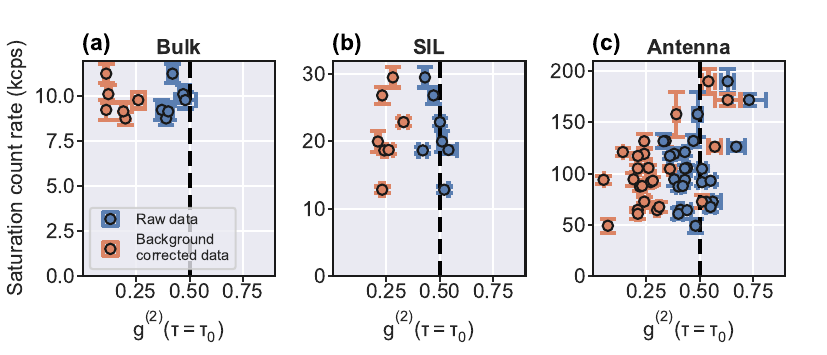}
\caption[]{
\textbf{Saturation count rates and autocorrelation drops at zero time delay.}
\textbf{(a)} Saturation count rate plotted over the fitted value of the autocorrelation measurement dip at zero time delay for representative V2 emitters in the bulk before (blue) and after (orange) background correction.
The same depiction is shown in \textbf{(b)} for all checked V2 emitters in the SIL sample and in \textbf{(c)} for all measured V2 emitters in the antenna sample.
The error bars in x (y) are the standard deviations from the corresponding saturation (autocorrelation) fits. Black, dashed lines indicate the single-emitter threshold of g$^{(2)}(0) = 0.5$ 
} 
\label{SuppFigG2}
\end{figure}

\noindent Figure \ref{SuppFigG2} shows the fitted saturation count rate over the fitted autocorrelation at zero time delay for all V2 emitters of our room temperature study before (blue) and after (orange) applying a background correction for the autocorrelation fit.
This correction is based on the fitted background from the corresponding saturation study.
For the bulk emitters in Figure \ref{SuppFigG2}(a), only six representative emitters of the 40 emitters from the statistics where investigated since autocorrelation measurements take a lot of time to yield a reasonable signal to noise ratio at the low photon count levels of bulk emitters.
Considering only the 14 antenna emitters with an anti-bunching drop below 0.5 without background correction yields an average enhancement of $\times \left(9.4 \pm 2.6 \right)$ and a maximum enhancement of $\times \left(12.5 \pm 1.8 \right)$, which is very similar to the statistics including all emitters after background correction as presented in the main manuscript.

\section{PLE categories of the antenna emitter} \label{sec:PLEcategories}
\begin{figure}[tb]
\centering
   \includegraphics[width=1\textwidth]{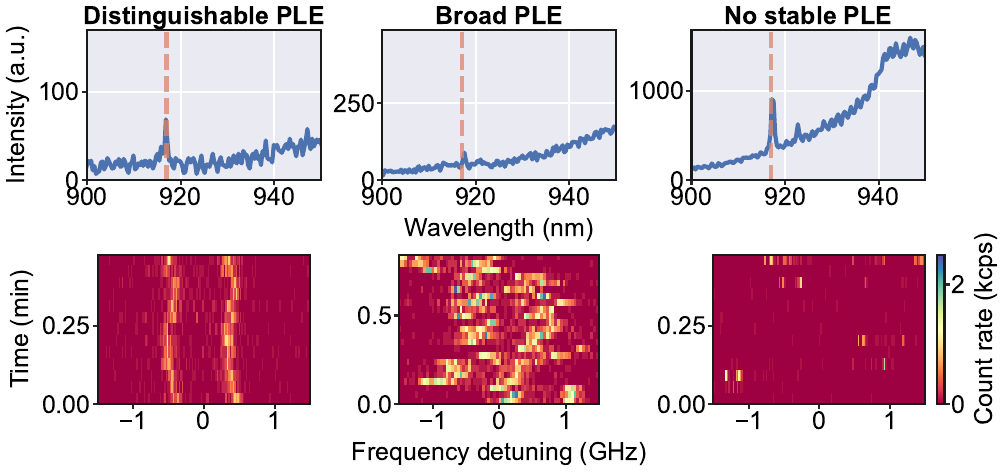}
\caption[]{
\textbf{Low-temperature emission spectra and PLE of V2 centers in the antenna sample.}
\textbf{(Upper row):} Emission spectra under off-resonant excitation (blue).
Orange, dashed lines indicate the position of the V2 ZPL for bulk samples. 
\textbf{(Lower row):} PLE for the spots with respective emission spectrum in the upper row.
}
\label{SuppFigPLEExample}
\end{figure}

\noindent To find emitters at low temperature for measuring PLE, we take emission spectra of bright spots on our sample and investigate spots with a clear peak around the V2 ZPL of $\sim \SI{917}{\nano\meter}$.
We consider all spots that show an emission peak in the range of \SIrange{916.2}{918.2}{\nano\meter} for subsequent PLE measurements, since this range was found to be accessible by tuning the grating of our resonant laser.
Our PLE measurements show emitters that can be assigned to one of three categories: 1.) emitters that show two, clearly distinguished absorption lines, 2.) emitters with a very broad absorption linewidth such that the two lines can hardly be distinguished, and 3.) emitters where only occasional blinking but no stable lines are observed at all.
Figure \ref{SuppFigPLEExample} shows the spectra (upper row) and a series of PLE lines (lower row) for three exemplary emitters of these categories.

\newpage
\section{Brightness of antenna emitters at resonant excitation} \label{sec:PLEsaturation}
\noindent As we discuss in the main manuscript, we extract the enhancement of the structure under resonant excitation through power dependent double-laser PLE measurements for 3 emitters in the bulk, 3 in the SIL, and 10 emitters in the antenna sample.
For all emitters, we measure 10 PLE lines for each excitation power and stop the measurement at the excitation power where we ionize the emitter regularly after 1-2 lines.
At next, we fit each single PLE line using a Gaussian and consider only fit results with a r$^2$-value of 0.8 or higher to discard spectrally unstable lines or lines that ionize already during the PLE measurement.
For each emitter, we plot the average fitted amplitude as a function of the excitation power and fit the data with a saturation curve, as for our room temperature study.
Here, we neglect the linear term in the saturation that accounts for the background, since background counts are highly suppressed at resonant excitation. 

\begin{figure}[tb]
\centering
   \includegraphics[width=\textwidth]{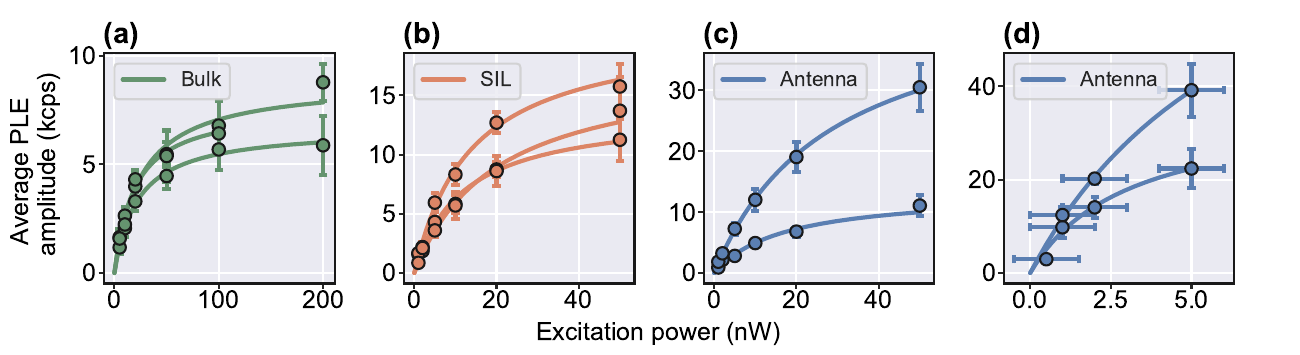}
\caption[]{
\textbf{Saturation count rates under resonant excitation.}
Count rates under resonant excitation extracted from the amplitudes of Gaussian fits on the PLE lines for 3 emitters in the bulk \textbf{(a)}, 3 in the SIL \textbf{(b)}, and 4 emitters in the antenna sample \textbf{(c)} and \textbf{(d)}.
Solid lines are saturation fits to the data.
The y-error bars in the data points are the standard deviation of the amplitude fits of the PLE lines.
Due to the significantly different saturation powers, the results from the four emitters of the antenna in (c) and (d) are plotted separately.}
\label{SuppFigSaturation}
\end{figure}

\begin{figure}[tb]
\centering
   \includegraphics[width=0.75\textwidth]{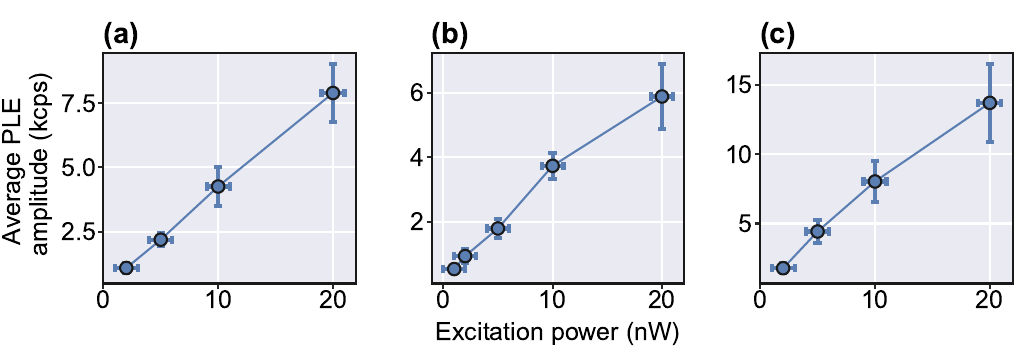}
\caption[]{
\textbf{Excitation power dependent count rates without saturation.}
Count rate as a function of resonant excitation power for 3 different emitters \textbf{(a)} - \textbf{(c)} in the antenna sample.
The solid line is a guide to the eye that underlines that for all three cases no saturation is visible yet.
} 
\label{SuppFigSaturationIssues}
\end{figure}

The results are depicted in Figure \ref{SuppFigSaturation} for the 3 bulk and SIL emitters as well as for 4 emitters from the antenna sample where the saturation power could be confidently extracted.
The emitters from the antenna sample are shown in the two plots \ref{SuppFigSaturation}(c) and (d), since they show a big difference in the maximum excitation power just before ionization.
We believe that this is a result of different power levels that emitters close to (away from) the membrane center experience at the same excitation power due to the cavity resonance being closer to (further from) \SI{917}{\nano\meter}.
The average saturation count rates for the three cases are $\SI{7.8 \pm 0.9}{\kilo\countpersec}$ (bulk), $\SI{17.4 \pm 2.9}{\kilo\countpersec}$ (SIL) and $\SI{48.2 \pm 30.2}{\kilo\countpersec}$ (antenna).
Note, that especially for the antenna emitters the saturation count values fluctuate strongly between different emitters. The resulting enhancement factors for the SIL and antenna are $\times \left( \SI{2.23 \pm 0.45}{} \right)$ and $\times \left( \SI{6.18 \pm 3.94}{} \right)$, respectively.
Apart from the emitter placement that plays a role for the enhancement in the antenna, we can attribute the fluctuation to missing data points in the saturated regime.
We could not record these points due to earlier ionization and even find emitters in the antenna where no saturation at all can be observed in the non-ionizing power range.
Three examples of this case are shown in Figure \ref{SuppFigSaturationIssues}.  

\newpage

\bibliography{references}